\documentclass{jpsj2}
\usepackage{bm}

\renewcommand{\d}{\text{d}}

\title{%
Diffusion Coefficient and Mobility of a Brownian Particle in a Tilted Periodic Potential
}

\author{%
Kazuo \textsc{Sasaki}\thanks{
E-mail address: sasaki@camp.apph.tohoku.ac.jp.} and Satoshi \textsc{Amari}
}

\inst{%
Department of Applied Physics, Tohoku University, 
Aoba-yama, Sendai 980-8579
}

\recdate{\today}

\abst{%
The Brownian motion of a particle in a one-dimensional periodic potential
subjected to a uniform external force $F$ is studied.
Using the formula for the diffusion coefficient $D$ obtained by other authors and an alternative one derived from the Fokker-Planck equation
in the present work, 
$D$ is compared with the differential mobility
$\mu = \text{d}v/\text{d}F$ where $v$ is the average velocity of the particle.
Analytical and numerical calculations indicate that inequality 
$D \ge \mu k_\text{B}T$, with $k_\text{B}$ the Boltzmann constant 
and $T$ the temperature, holds if the periodic potential is symmetric,
while it is violated for asymmetric potentials when $F$ is small but nonzero.
}

\kword{%
Brownian motion, diffusion coefficient, mobility, Fokker-Planck equation
}

\begin{document}

\sloppy
\maketitle

\section{\label{sec:intro}Introduction}

The response of a system in thermal equilibrium to an external disturbance
has close relation to fluctuations produced spontaneously in the system
in the absence the disturbance.
This relation can be formulated as the fluctuation-dissipation theorem.\cite{kubo91, zwanzig01}
The Einstein relation $D = \mu_0 k_\text{B}T$ is a famous example,
where $D$ is the diffusion coefficient and $\mu_0$ is the mobility of a
Brownian particle, $k_\text{B}$ is the Boltzmann constant, and $T$ is
the temperature.
In this example, $D$ measures the fluctuation of the particle position or velocity $v$
and $\mu_0 = \lim_{F \to 0}v/F$ represents the response of the
particle velocity to a small external force $F$.

For systems far from thermal equilibrium, 
any particular relation between $D$ and $\mu_0$ is expected,
because we do not know general laws like the fluctuation-dissipation theorem
for such systems.
However, recent investigations\cite{reimann01, sasaki03, harada04}
into certain one-dimensional systems in nonequilibrium steady states suggest
that inequality $D \ge \mu k_\text{B}T$ with $\mu = \text{d}v/\text{d}F$ being 
the \textit{differential mobility\/} may hold in these systems:
numerical data show that $D$ is greater than $\mu k_\text{B}T$ 
for a Brownian particle moving in sinusoidal
 potentials\cite{reimann01}, 
 for flushing ratchets\cite{sasaki03} and for rocking ratchets\cite{harada04}.
 Is there any rule that tells under what conditions inequality $D \ge \mu k_\text{B}T$
 holds?
 Finding such a rule, if exists,  would provide an important insight into
 understanding the behavior of  nonequilibrium systems.
 
 The purpose of the present paper is to figure out whether inequality
$D \ge \mu k_\text{B}T$ holds generally in the system of a Brownian particle 
moving in a one-dimensional periodic potential subjected to a uniform
external force.
This system is one of the simplest systems that exhibit nonequilibrium steady states,
and convenient formulas for calculating $D$ and $\mu$ are 
known\cite{reimann01, hayashi04}.
From analytical and numerical investigations based on these formulas
and the one we derive from the steady-state solution to the Fokker-Planck 
equation in the present work, 
we find that this inequality is likely to be valid for any symmetric potentials
whereas it is violated for small external forces if the potential is asymmetric.

\section{Formulas}
\label{sec:formulas}

We shall investigate the overdamped motion of a Brownian particle moving along the
$x$ axis under the influence of a periodic potential $V(x)$ 
of period $l$ and a uniform external force $F$.
The total potential $U(x)$ for the particle is given by
\begin{equation}
\label{eq:Ux}
	U(x) = V(x) - Fx, \quad V(x + l) = V(x).
\end{equation}
In what follows periodic functions $I_\pm(x) = I_\pm(x + l)$ defined by
\begin{equation}
\label{eq:Ipm}
	I_\pm(x) = \frac{1}{l}\int_0^l \text{e}^{\pm\beta U(x) \mp \beta U(x \mp y)}\,\d y
\end{equation}
play important roles,
where $\beta =1/k_\text{B}T$.
The average of a periodic function $f(x)$ of period $l$ over the period
will be denoted 
by $\langle{f}\rangle$:
\begin{equation}
\label{eq:Iav}
	\langle{f}\rangle = \frac{1}{l}\int_0^l f(x)\,\d x,
	\quad f(x + l) = f(x).
\end{equation}
The ``normalized'' functions
\begin{equation}
\label{eq:Jpm}
	J_\pm(x) = I_\pm(x)/\langle{I_\pm}\rangle,
\end{equation}
which satisfy $\langle{J_\pm}\rangle  = 1$, are also of use.

It was shown by Stratonovich\cite{strat58} that the average velocity $v$ of 
the particle can be calculated by the formula\cite{risken89, reimann01}
\begin{equation}
\label{eq:Strat}
	v = D_0(1 - \text{e}^{-\beta Fl})/l\langle{I_\pm}\rangle,
\end{equation}
where  $D_0$ is the diffusion coefficient of a freely moving Brownian particle
($V = F = 0$) and it is related with the frictional coefficient $\zeta$ of
the particle through $D_0 = k_\text{B}T/\zeta$.
It is noted that $\langle{I_{+}}\rangle = \langle{I_{-}}\rangle$. 
The differential mobility $\mu = \text{d}v/\text{d}F$ can be
calculated by differentiating eq.~(\ref{eq:Strat}) with respect to $F$.
The result can be expressed in a succinct form:\cite{hayashi04}
\begin{equation}
\label{eq:mob}
	\mu 
	= D_0\langle{J_{+}J_{-}}\rangle/k_\text{B}T.
\end{equation}
The formula for $D$ in the presence of both $V(x)$ and $F$ was
derived recently by Reimann et al.:\cite{reimann01}
\begin{equation}
\label{eq:Reim}
	D = D_0\langle{J_\pm J_{+}J_{-}}\rangle.
\end{equation}
Note that $\langle{J_{+}^2J_{-}}\rangle$ is equal to 
$\langle{J_{+}J_{-}^2}\rangle$.
Reimann et al.\cite{reimann01} derived this formula by considering the 
moments of first passage time.
Later, Hayashi and Sasa\cite{hayashi04} obtained the same result by considering
the system with an additional potential that varies much slowly than 
the original periodic potential $V(x)$. 

If the periodic potential $V(x)$ and the external force $F$ are given,
the diffusion coefficient $D$ and the differential mobility $\mu$ can be
figured out by carrying out the two-dimensional integrals involved in 
eqs.~(\ref{eq:Reim}) and (\ref{eq:mob});
from the results we find whether or not $D$ is larger than $\mu k_\text{B}T$.
Nevertheless, an alternative formula may be useful in studying the sign
of $D - \mu k_\text{B}T$.
From the steady-state solution to the Fokker-Planck equation 
we can derive (see the appendix) the formula
\begin{equation}
\label{eq:Dm}
	D - \mu k_BT = vl\langle(J_{+} - 1)K_{-}\rangle,
\end{equation}
where periodic functions $K_{\pm}(x)$ of period $l$ are defined by
\begin{equation}
\label{eq:Kdef}
	K_{\pm}(x) = \frac{1}{l}\int_0^x[J_{\pm}(y) - 1]\,\d y.
\end{equation}
Because the sign of $v$ is the same as that of $F$ as evident from 
eq.~(\ref{eq:Strat}),
formula (\ref{eq:Dm}) indicates that
$D > \mu k_\text{B}T$ if the sign of 
\begin{equation}
\label{eq:sdef}
	s = \langle{(J_{+} - 1)K_{-}}\rangle
\end{equation}
is the same as that of $F$.
In analytic investigations, evaluation of eq.~(\ref{eq:sdef}) is usually
much easier than calculating eqs.~(\ref{eq:mob}) and (\ref{eq:Reim})
and then subtracting one from the other.
By contrast, it is better to use eqs.~(\ref{eq:mob}) and (\ref{eq:Reim})
in numerical calculations, because the evaluation of the three-dimensional integral involved in
eq.~(\ref{eq:sdef}) is time consuming.

\section{Example}
\label{sec:example}

In this section we present the numerical results for the diffusion coefficient $D$ and the differential mobility $\mu$ obtained from formulas (\ref{eq:Reim}) and (\ref{eq:mob}), respectively, with a particular choice of potential:
\begin{equation}
\label{eq:V1}
	V(x) = A[\sin(2\pi x/l) - \lambda\sin(4\pi x/l)],
\end{equation}
where $A > 0$ and $\lambda$ are parameters.
This potential is symmetric if $\lambda = 0$ or $\lambda = \infty$,
 and asymmetric otherwise.
The potential height $W$, defined as the difference between the maximum and minimum values of $V$, is given by
\begin{equation}
\label{eq:W1}
	W = 2(1 - 2\lambda c)\sqrt{1 - c^2}\,A,
\end{equation}
where $c$ is defined by 
\begin{equation}
\label{eq:cdef}
	c = (1 - \sqrt{1 + 32\lambda^2})/8\lambda.
\end{equation}
Note that this potential has a single minimum and a single maximum in each period
if $0 \le |\lambda| < 1/2$,
while it has an extra pair of local minimum and maximum if $1/2 < |\lambda| < \infty$.

\begin{figure}[t]
	\begin{center}
		\includegraphics[width=7.5cm, clip]{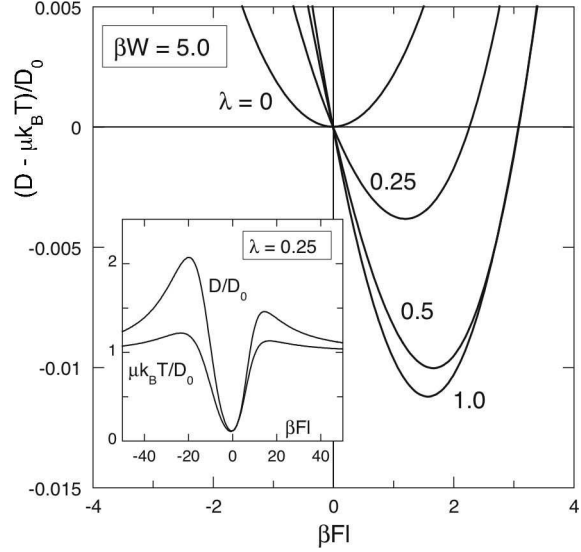}
	\end{center}
	\caption{The difference $D - \mu k_\text{B}T$ in units of $D_0$ as a
	function of the external force $F$ in the dimensionless form ($\beta F l$)
	for the system with the potential given by eq.~(\ref{eq:V1}). 
	The potential height is chosen as $W = 5.0\,k_\text{B}T$,
	and  the results for different values of parameter $\lambda$ are shown.
	The inset presents the dependence of $D/D_0$ and $\mu k_\text{B}T/D_0$	
	on $\beta Fl$ in the case that $\beta W = 5.0$ and $\lambda = 0.25$.
	}
	\label{fig:dm}
\end{figure}

The inset of Fig.~\ref{fig:dm} shows the dependence of $D$ and $\mu k_\text{B}T$
on the external field $F$ in the case that $\beta W = 5.0$ and $\lambda = 0.25$.
It appears that $D$ is always larger than $\mu k_\text{B}T$.
However, closer inspection reveals that $D$ is smaller than $\mu k_\text{B}T$ 
in a certain range of $F$ near $F = 0$:
See Fig.~\ref{fig:dm}, where the difference $D - \mu k_\text{B}T$ is plotted 
against $F$ in expanded scales for several values of $\lambda$ with $\beta W = 5.0$;
the results for negative values of $\lambda$ is obtained from the corresponding
results for $-\lambda$ (which is now positive) by changing the sign of $F$, as
the symmetry property indicates.
In the case of symmetric potential ($\lambda = 0$) we observe that 
$D \ge \mu k_\text{B}T$ (the equality holds when $F = 0$).

\begin{figure}[t]
	\begin{center}
		\includegraphics[width=7cm, clip]{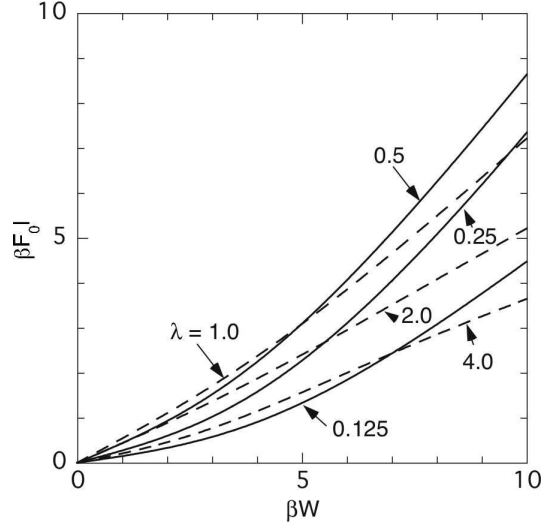}
	\end{center}
	\caption{The upper bound $F_0$ of the interval for external force in which inequality
	$D < \mu k_\text{B}T$ holds is plotted as a function of the potential height $W$
	for the system with the potential given by eq.~(\ref{eq:V1}).
	The results for different choices of parameter $\lambda$ are shown.
	}
	\label{fig:f0}
\end{figure}

For potentials with positive (negative) $\lambda$, 
we find that $D < \mu k_\text{B}T$ in a range  $0 < F < F_0$ ($F_0 < F < 0$)
of $F$ where the upper (lower) bound $F_0$ depends on $\lambda$, $W$ and $\beta$.
Figure~\ref{fig:f0} shows the dependence of $\beta F_0l$ on $\beta W$ for
several values of positive $\lambda$.
One sees that $\beta F_0l$ is a monotonically increasing function of $\beta W$.
If $\lambda \le 1/2$ (the solid lines in Fig.~\ref{fig:f0}),
the value of $\beta F_0l$ for a fixed $\beta W$ decreases with
decreasing $\lambda$ and becomes zero as $\lambda = 0$ is approached.
By contrast, $\beta F_0l$ decreases with increasing $\lambda$ when $\lambda \ge 1$
(the dashed lines in Fig.~\ref{fig:f0}).
These behaviors may be summarized that as the potential becomes symmetric,
the range of $F$ in which inequality $D < \mu k_\text{B}T$ holds shrinks to zero.

\section{Conjectures}
\label{sec:conjectures}

\begin{figure}[t]
	\begin{center}
		\includegraphics[width=7cm, clip]{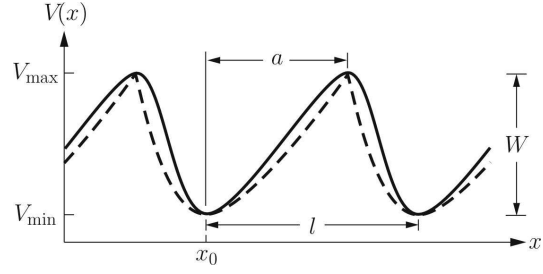}
	\end{center}
	\caption{Two examples of periodic potential $V(x)$ of period $l$ 
	that has a single minimum and a single maximum  in a period 
	are schematically shown.
	The one represented by the solid line has rounded peaks at its maxima
	and rounded valleys at its minima, while the one represented by the
	dashed line has cusps at its maxima.
	The location of a minimum is $x_0$, and the the distance from this 
	minimum to the adjacent maximum on the right is $a$.
	The potential height $W$ is defined as the difference between the 
	maximum and minimum values ($V_\text{max}$ and $V_\text{min}$) of $V$.
	}
	\label{fig:vx}
\end{figure}

We have calculated $D$ and $\mu$ numerically for various periodic potentials $V(x)$ 
in  addition to the one described in the preceding section; 
some of the results will be presented in the following section.
We have also carried out analytic study on the sign of $D - \mu k_\text{B}T$
in several limiting cases, which will be discussed in the next section.
From the results of these investigations,
we have been lead to postulate the following conjectures.
\begin{itemize}
\item[(i)]
Inequality $D \ge \mu k_\text{B}T$ holds for arbitrary \textit{symmetric} potentials $V$.
\item[(ii)]
Suppose that the potential is \textit{asymmetric} and 
has a single minimum and a single maximum in each period.
Let $a$ be the distance from a minimum to the adjacent maximum 
on the right (see Fig.~\ref{fig:vx}).
Then, 
we have $D < \mu k_\text{B}T$ for $0 < F < F_0$ ($F_0 < F < 0$)
and $D \ge \mu k_\text{B}T$ outside this interval of $F$
if $a > l/2$ ($a < l/2$),
where $F_0$  is a positive (negative) constant that depends on potential $V(x)$
and temperature $T$.
\end{itemize}
Note that in the example considered in the preceding section 
distance $a$ is given by 
\begin{equation}
\label{eq:aV1}
	a = (l/\pi)\arccos c,
\end{equation}
where $c$ is defined by eq.~(\ref{eq:cdef}), 
and hence condition $a > l/2$ corresponds to $\lambda > 0$.
Therefore, the results shown in Fig.~\ref{fig:dm} are consistent with 
these conjectures.

\section{Evidence}
\label{sec:evidence}

The conjectures stated in the preceding section are based on the analyses presented
in this section.
We first describe the analytical investigations, in several limiting cases,
into the sing of $D - \mu k_\text{B}T$ using formula~(\ref{eq:Dm}).
Then, considering the results of these investigations and supplementary 
numerical calculations, we will argue for the validity of the conjectures.

\subsection{Small external force}
\label{sec:smallF}

The first limiting case we study is the case of small external force represented by 
condition $\beta|F|l \ll 1$.
In this case the factor $s$ defined by eq.~(\ref{eq:sdef}) may be expanded
in powers of $\beta Fl$ as
\begin{equation}
\label{eq:sexp}
	s  = s_0 + s_1 \beta Fl + s_2(\beta Fl)^2 + \dots .
\end{equation}
In order to express the expansion coefficients $s_0$, $s_1$ and so on concisely,
we introduce periodic functions (of period $l$) $\psi_{\pm}(x)$ 
and $\chi_{\pm}(x)$ by
\begin{equation}
\label{eq:psidef}
	\psi_{\pm}(x) = \text{e}^{\pm\beta V(x)}/
		\langle{\text{e}^{\pm\beta V}}\rangle
\end{equation}
and
\begin{equation}
\label{eq:chidef}
	\chi_{\pm}(x) = \frac{1}{l}\int_0^x [\psi_{\pm}(y)  - 1]\,\d y.
\end{equation}
It is not difficult to see that periodic functions $I_\pm(x)$ 
defined by eq.~(\ref{eq:Ipm}) can be expressed as
\begin{align}
	I_\pm(x) &= \langle{\text{e}^{\mp\beta V}}\rangle \text{e}^{\pm\beta V(x)}
		\bigl\{1 - \beta Fl
		\left[1/2 \mp \chi_\mp(x) \pm \langle{\chi_\mp}\rangle\right] 
		\nonumber\\
		&\quad{}+ O[(\beta Fl)^2]\bigr\}.
		\label{eq:Iexp}
\end{align}
Therefore, the normalized functions $J_\pm(x)$ defined by eq.~(\ref{eq:Jpm}) are
given by
\begin{align}
	J_\pm(x) &= \psi_\pm(x)\bigl\{1 \pm \beta Fl
		\left[\chi_\mp(x) - \langle{\psi_\pm\chi_\mp}\rangle\right]
		\nonumber\\
		&\quad{}+ O[(\beta Fl)^2]\bigr\},
		\label{eq:Jexp}
\end{align}
from which the following expression for $K_\pm(x)$ defined by
eq.~(\ref{eq:Kdef}) is obtained:
\begin{align}
	K_\pm(x)  &= \chi_\pm(x) \pm \beta F
		\int_0^x \psi_\pm(y)[\chi_\mp(y) 
		- \langle{\psi_\pm\chi_\mp}\rangle]\,\d y
		\nonumber\\
		&\quad{}+ O[(\beta Fl)^2].
		\label{eq:Kexp}
\end{align}
Substituting eqs.~(\ref{eq:Jexp}) and (\ref{eq:Kexp}) into eq.~(\ref{eq:sdef}),
one finds
\begin{equation}
\label{eq:s0}
	s_0 = \langle{(\psi_{+} - 1)\chi_{-}}\rangle
\end{equation}
and
\begin{equation}
\label{eq:s1}
	s_1 = \langle{\psi_{+}\chi_{-}^2}\rangle - \langle{\psi_{+}\chi_{-}}\rangle^2
		+ \langle{\psi_{-}\chi_{+}^2}\rangle - \langle{\psi_{-}\chi_{+}}\rangle^2.
\end{equation}

It is worth noting that coefficient $s_1$ cannot be negative:
\begin{equation}
\label{eq:s1pos}
	s_1 \ge 0,
\end{equation}
where the equality holds only in the trivial case of a constant potential $V$.
This property comes from the Schwarz inequality
\begin{equation}
\label{eq:Schw}
	\langle{\psi_\pm}\rangle\langle{\psi_{\pm}\chi_{\mp}^2}\rangle
	\ge \langle{\psi_{\pm}\chi_{\mp}}\rangle^2
\end{equation}
and identity $\langle{\psi_\pm}\rangle = 1$ resulting from the definition 
(\ref{eq:psidef}) of $\psi_\pm$;
the equality in eq.~(\ref{eq:Schw}) holds if and only if $\chi_\pm$ is a constant
(i.e., $V$ is a constant).

By contrast, 
the leading term $s_0$ in expansion~(\ref{eq:sexp}) can be positive or negative.
However, in the case of symmetric potential, i.e., 
if there exists a constant $\alpha$ such that $V(x) = V(2\alpha - x)$ holds for any $x$,
we have $s_0 = 0$.
The reason is the following: for such a symmetric potential,
 $\psi_\pm(x)$ is symmetric and $\chi_\mp(x)$ is antisymmetric about $x = \alpha$,
hence we obtain $\langle{\psi_\pm\chi_\mp}\rangle = 0$ and
$\langle{\chi_\mp}\rangle = 0$, which imply $s_0 = 0$.
This fact and inequality~(\ref{eq:s1pos}) indicate that 
inequality $D \ge \mu k_\text{B}T$ holds for
any symmetric potentials as long as $F$ is small,
which supports conjecture (i) stated in the preceding section.

In the case of asymmetric potential, 
it is expected that $s_0 \not= 0$.
Then, what property of $V$ determines the sign of $s_0$
(i.e., the sign of $D - \mu k_\text{B}T$ for small $F$)?
It seems difficult to answer this question for arbitrary potentials.
However, if we restrict our attention to a certain class of potentials,
we can, at least partly, answer the question.
Let us consider a potential that has only one minimum and one maximum in 
a period, as shown in Fig.~\ref{fig:vx}.
Let $x_0$ be the location of a minimum, 
$a$ be the distance from a minimum to the adjacent maximum on the right,
$W$ be the potential height defined as the difference between the maximum
and minimum values of $V$.
The potential $V$ may have a rounded peak at its maximum and a rounded valley at its
minimum as shown by the solid line in Fig.~\ref{fig:vx}.
It may have a cusp at its maximum (the dashed line in Fig.~\ref{fig:vx}),
or at its minimum, or at both. 
We shall analyze the sign of $s_0$ in the limiting cases of large potential height
($\beta W \gg 1$) and small potential height ($\beta W \ll 1$).

Let us consider the case of large potential height, $\beta W \gg 1$.
In order to make the analysis simple, we assume that the origin of the $x$ axis
is chosen in such a way that condition $0 < x_0 < x_0 + a < l$ is satisfied.
In evaluating $s_0$ given by eq.~(\ref{eq:s0}),
it is noted that function $\psi_{+}(x)$ has a sharp peak at $x = x_0 + a$
and vanishes rapidly as one moves away from the peak.
Therefore $s_0$ can be approximated by
\begin{equation}
\label{eq:s0a}
	s_0 \simeq \langle{\psi_{+}}\rangle\chi_{-}(x_0 + a) - \langle{\chi_{-}}\rangle,
\end{equation}
since $\chi_{-}(x)$ does not vary rapidly in the vicinity of $x = x_0 + a$
as we shall see in a moment.
Function $\chi_{-}(x)$ defined by eq.~(\ref{eq:chidef}) is the sum of
\begin{equation}
\label{eq:sigma}
	\sigma(x) = \frac{1}{l}\int_0^x \psi_{-}(y)\,\d y
\end{equation}
and $-x/l$.
Since the integrand $\psi_{-}(y)$ in eq.~(\ref{eq:sigma}) is practically zero
except a narrow region around the sharp peak at $y = x_0$,
function $\sigma(x)$ behaves like a step function:
as $x$ is increased from zero to $l$, $\sigma(x)$ increases rapidly from zero to
unity around $x = x_0$.
Therefore, $\chi_{-}(x)$ is well approximated by $\chi_{-}(x) \simeq 1 - x/l$ near
$x= x_0 + a$ and it does not change rapidly in the vicinity of $x= x_0 + a$.
We also find that $\langle{\chi_{-}}\rangle = 1/2  - x_0/l$ if the small 
correction of order $1/\beta W$ is neglected.
From these arguments and identity $\langle{\psi_{+}}\rangle = 1$
we obtain
\begin{equation}
\label{eq:s0b}
	s_0 \simeq \frac{1}{2} - \frac{a}{l}.
\end{equation}
This expression for $s_0$ reveals that the sign of $s_0$ is determined by
whether the location of the top of the potential hill between
a pair of neighboring valleys is closer to the left valley ($a/l < 1/2$) or 
the right one ($a/l > 1/2$),
which supports conjecture (ii) in the preceding section.

Now we turn our attention to the case of small potential height, $\beta W \ll 1$.
It will be assumed that (an arbitrary constant is added to $V$ such that)
the maximum and minimum values of $V$ are of order $W$.
Then condition $\beta W \ll 1$ implies $|\beta V| \ll 1$.
Expanding $\psi_{\pm}(x)$ and $\chi_{\pm}(x)$ defined by eqs.~(\ref{eq:psidef})
and (\ref{eq:chidef})  in powers of $\beta V$,
and then substituting them into eq.~(\ref{eq:s0}), 
we obtain
\begin{equation}
\label{eq:s0c}
	s_0 = -\beta^3\langle (V^2 - \langle V^2\rangle) \mathcal{V} \rangle
		+ O[(\beta V)^4],
\end{equation}
where periodic function $\mathcal{V}(x)$ of period $l$ is defined by
\begin{equation}
\label{eq:defv}
	\mathcal{V}(x) = \frac{1}{l}\int_0^x[V(y) - \langle{V}\rangle]\,dy.
\end{equation}

Unlike the case of $\beta W \gg 1$, 
we have not been able to relate the sign of $s_0$ approximated by 
eq.~(\ref{eq:s0c}) to that of $l/2 - a$ for general asymmetric potentials.
Here we investigate the sign of $s_0$ for three examples of potential $V(x)$.
The first example is the one considered in Sec.~\ref{sec:example},
see eq.~(\ref{eq:V1}).
The second example is a piecewise-cubic function given by
\begin{equation}
\label{eq:V2}
	V(x) = A\{(x/l)^2 + \lambda (x/l)[1 - 4 (x/l)^2]\} \quad
	(|x| \le l/2),
\end{equation}
where $A$ and $\lambda$ are parameters;
$V(x)$ outside the range $|x| \le l/2$ is defined such that it is a periodic function
of period $l$.
We shall assume that $A > 0$ and $|\lambda| < 1/2$.
Then $V(x)$ has a cusp at its maximum, 
as the one represented by the dashed line in Fig.~\ref{fig:vx} does.
The distance $a$ from a minimum to the adjacent maximum on the right
is given by
\begin{equation}
\label{eq:a2}
	a = l\left(\frac{1}{2} + \frac{\lambda}{1 + \sqrt{1 + 12\lambda^2}}\right),
\end{equation}
and the potential height $W$ by
\begin{equation}
\label{eq:W2}
	W = A\left[\frac{2}{9} +
		\frac{1 + 12\lambda^2}{18(1 + \sqrt{1 + 12\lambda^2})}\right].
\end{equation}
The third example is a piecewise-linear (sawtooth) potential
\begin{equation}
\label{eq:V3}
	V(x) = \begin{cases}
		Wx/a & 0 \le x < a \\
		W(l - x)/(l - a) & a \le x < l,
	\end{cases}
\end{equation}
where $W$ and $a$ are positive parameters with restriction $0 <a < l$;
again, $V(x)$ outside the range $0 \le x < l$ is defined such that it is a periodic function
of period $l$.
Parameter $W$ represents the potential height,
and parameter $a$ corresponds to the distance from a minimum of  $V(x)$ and the
adjacent maximum on the right.

For each example, the leading term of $s_0$ given in eq.~(\ref{eq:s0c}) 
has been calculated.
The results are summarized in Table~\ref{tab:s0}.
In all the three examples the sign of $s_0$ is the same as that of $l/2 - a$
(remember that $l/2 > a$ if $\lambda < 0$ in the first two examples).
This observation is consistent with conjecture (ii).

\begin{table}
\caption{
	Approximate expression for $s_0$ given as the leading term in 
	eq.~(\ref{eq:s0c}) obtained for three examples of $V(x)$.
}
\vspace{5pt}
\label{tab:s0}
\begin{tabular}{cccc}
\hline
 & Example 1 & Example 2 & Example 3 \\
 \hline
 $V(x)$ & eq.~(\ref{eq:V1}) & eq.~(\ref{eq:V2}) & eq.~(\ref{eq:V3}) \\ [6pt]
 $s_0$ & 
 $
 -\frac{3\lambda(\beta A)^3}{16\pi}$ &
 $
 -\frac{\lambda(\beta A)^3}{1575}
 	(\frac{1}{12} + \frac{1}{11}\lambda^2)$ &
$
\frac{(\beta W)^3}{360}(1 - \frac{2a}{l})$ \\ [8pt]
\hline
\end{tabular}
\end{table}

 It is interesting to note that $s_0$ is of higher order in $\beta W$ than
 \begin{equation}
 \label{eq:s1b}
 	s_1 = 2\beta^2\left(\langle{\mathcal{V}^2}\rangle 
	 	- \langle{\mathcal{V}}\rangle^2\right)
	 		+ O[(\beta V)^3]
\end{equation}
in the case of small potential height. 
This fact implies that $s$ change its sign at small $\beta Fl$ when the latter is varied.
Let $F_0$ be the value of $F$ at which $s$ changes its sign, 
then one finds from eq.~(\ref{eq:sexp}) that
\begin{equation}
\label{eq:F0}
	\beta F_0 l \simeq -\frac{s_0}{s_1}
	\simeq \frac{\beta\langle (V^2 - \langle V^2 \rangle) \mathcal{V} \rangle}
		{2(\langle{\mathcal{V}^2}\rangle - \langle{\mathcal{V}}\rangle^2)},
\end{equation}
which is of order $\beta W$.
For the first example considered above, we obtain
\begin{equation}
\label{eq:F01}
	\beta F_0l \simeq \frac{3\pi\lambda}{4(1 + \lambda^2/4)}\beta A.
\end{equation}
This relation and eq.~(\ref{eq:W1}) explain the behavior of the graphs 
in Fig.~\ref{fig:f0} near the origin.

\subsection{Large external force}
\label{sec:largeF}

If the external force $F$ is large enough ($\beta|F|l \gg 1$),
the dominant contribution to the integral in eq.~(\ref{eq:Ipm}) defining $I_{\pm}(x)$
comes from the narrow region near $y = 0$ (if $F > 0$) or $y = l$ (if $F < 0$).
Therefore, $I_{\pm}(x)$ may be expanded as
\begin{equation}
\label{eq:iexpf}
	I_{\pm}(x) \simeq \frac{\text{e}^{\pm \beta V(x)}}{l}\int_0^\infty 
	\left[h_{\mp}(0) + yh_{\mp}'(0) + \dots \right]
	\text{e}^{-\beta Fy}\,dy
\end{equation}
if $F > 0$, 
where  $h_{\mp}(y) = \text{e}^{\mp\beta V(x \mp y)}$, and
$h_{\mp}'(y)$ is the derivative of $h_{\mp}(y)$.
A similar expansion in the case of $F < 0$ can be made.
Using these expansions, 
periodic functions $I_{\pm}(x)$, $J_{\pm}(x)$, and $K_{\pm}(x)$ are
expressed as the power series in $1/F$.
Substitution of $J_{+}(x)$ and $K_{-}(x)$ thus obtained into eq.~(\ref{eq:sdef}) yields
\begin{equation}
\label{eq:sF}
	s = \frac{1}{\beta Fl}\left[\frac{2\langle{(V')^2}\rangle}{F^2} + 
	\frac{5\langle{(V')^3}\rangle}{F^3} + O\left(\frac{1}{F^4}\right)\right],
\end{equation}
where $V'(x)$ is the derivative of potential $V(x)$.
This expression is valid both for $F >0 $ and for $F < 0$.
The leading term of $s$ given by eq.~(\ref{eq:sF}) has the same sign as that of $F$
and hence inequality $D > \mu k_\text{B}T$ holds if $|F|$ is large enough.

\subsection{Small potential height}
\label{sec:smallW}

The last limiting case we study is the limit of small potential height;
the strength of the external force $F$ can be arbitrary.
In this case we find it convenient to express the potential $V(x)$ in the 
Fourier series as
\begin{equation}
\label{eq:fourier}
	V(x) = \sum_{n = -\infty}^\infty \hat V_n \text{e}^{ik_nx}, 
	\quad
	k_n = \frac{2\pi}{l}n.
\end{equation}
In the integrand of eq.~(\ref{eq:Ipm}), 
factor $e^{\mp\beta V(x \mp y)}$ is expanded in powers of $\beta V(x \mp y)$
and then eq.~(\ref{eq:fourier}) is substituted to carry out the integral.
Once $I_{\pm}(x)$ are obtained in this way,
it is straightforward to calculate $J_{\pm}(x)$ and $K_{\pm}(x)$.
Substituting the resulting expressions for $J_{+}(x)$ and $K_{-}(x)$ 
into eq.~(\ref{eq:sdef}), we have
\begin{equation}
\label{eq:sw}
	s = \sum_{n = 1}^\infty\frac{4\beta Fk_n^2|\beta \hat V_n|^2}
		{l[(\beta F)^2 + k_n^2]^2} + O[(\beta V)^3].
\end{equation}
The sign of the leading term in this expression for $s$ is the same as that of $F$, 
and hence  inequality $D > \mu k_\text{B}T$ holds if $\beta W$ is small enough.

It is noted that in the limit of small $\beta Fl$ the leading term in eq.~(\ref{eq:sw})
approaches to
\begin{equation}
\label{eq:swF}
	4\beta Fl\sum_{n = 1}^\infty\frac{|\beta \hat V_n|^2}{(lk_n)^2}
	= 2\beta^2(\langle\mathcal{V}^2\rangle 
		- \langle\mathcal{V}\rangle^2)\beta Fl .
\end{equation}
This expression agrees with the leading term of eq.~(\ref{eq:s1b}) multiplied by
$\beta Fl$.
This is expected from the consistency of the analysis.
Similarly, the term of order $(\beta V)^3$ in eq.~(\ref{eq:sw}) should
converge to the first term of $s_0$ given in eq.~(\ref{eq:s0b}) in the limit 
$F \to 0$, which we have not checked.
In the opposite limit, $\beta |F|l \gg 1$, the leading term in eq.~(\ref{eq:sw})
converges to the leading term in eq.~(\ref{eq:sF}),
because $\sum_n k_n^2|\hat V_n|^2 = \langle{(V')^2}\rangle$.

\subsection{Symmetric potentials}
\label{sec:symmetric}

Here, we consider the case of symmetric potential
and argue for the validity of conjecture (i).
In this case, $s$ is an odd function of $F$ 
($D - \mu k_\text{B}T$ is an even function of $F$), 
and therefore we need to examine the sign of $s$ only for $F \ge 0$.
Remember that $s > 0$ is equivalent to $D > \mu k_\text{B}T$ when $F > 0$.
It has been shown that 
\begin{equation}
\label{eq:ssym1}
	s \simeq s_1\beta Fl
\end{equation}
with $s_1 > 0$ for small $\beta Fl$ (\S\ref{sec:smallF}) and 
$s \simeq 2\langle(V')^2\rangle/\beta F^3l$ for large $\beta Fl$ (\S\ref{sec:largeF}).
Hence, it is concluded that inequality $D \ge \mu k_\text{B}T$ 
holds (the equality holds when $F = 0$) in these two extremes.
Furthermore, this inequality has been found to be valid in the entire range of $F$
if the potential height is small compared to the temperature (\S\ref{sec:smallW}). 

In order to assert the validity of conjecture (i), we have to demonstrate
that $s > 0$ for intermediate values of $\beta Fl$ when $\beta W$ is not small.
For this purpose, numerical calculations of $s = (D - \mu k_\text{B}T)/vl$ are
carried out using formula
\begin{equation}
\label{eq:sdef1}
	s = \frac{\langle{I_{+}^2I_{-}}\rangle 
	- \langle{I_{+}}\rangle\langle{I_{+}I_{-}}\rangle}
	{\langle{I_{+}}\rangle^2(1 - \text{e}^{-\beta Fl})}
\end{equation}
obtained from~(\ref{eq:Strat}), (\ref{eq:mob}) and (\ref{eq:Reim});
as remarked earlier, this method of evaluating $s$ is more convenient for
numerical calculations than using formula~(\ref{eq:sdef}).
Symmetric potentials $V(x)$ of the following type are examined:
\begin{equation}
\label{eq:vsym}
	V(x) = A\sum_{n = 1}^N c_n\cos(2n\pi x/l),
\end{equation}
where $N$ is a positive integer, $c_n$ are arbitrary coefficients,
and the overall factor $A$ is determined such that the potential height is $W$
for given values of $W$ and $c_1$, $c_2$, \dots, $c_N$.

\begin{figure}[t]
	\begin{center}
		\includegraphics[width=7.5cm, clip]{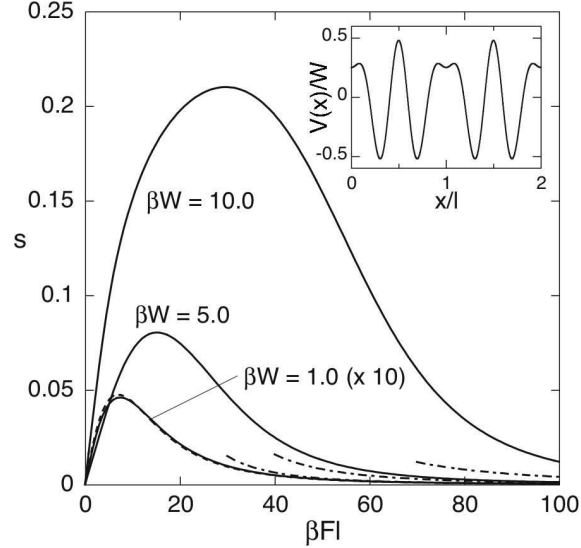}
	\end{center}
	\caption{Dependence of $s$ on  $\beta Fl$ numerically obtained for 
	a symmetric potential~(\ref{eq:vsym}) with $N = 3$ and $c_1 = 0.265947$, 
	$c_2 = 0.823433$, $c_3 = -0.522984$; the inset depicts the potential function.
	The results for different choices of potential height $W$ are shown.
	The dashed line represents the analytic expression~(\ref{eq:sw}) valid
	for small $\beta W$, 
	and the dash-dotted lines indicate the leading term in eq.~(\ref{eq:sF})
	for large $\beta Fl$.
	It is remarked that the graphs of 10 times $s$ instead of $s$ 
	are plotted for $\beta W = 1.0$.
	}
	\label{fig:sym3s}
\end{figure}

Figure~\ref{fig:sym3s} shows the numerical results for a potential with 
an arbitrarily chosen set of coefficients $\{c_n\}$ in the case of $N = 3$.
Here, $s$ is plotted as a function of $\beta Fl$ for several values of $\beta W$.
As $F$ is increased from zero, $s$ starts to increase linearly in $F$ as 
eq.~(\ref{eq:ssym1}) predicts and it continues to increase until it reaches 
a maximum value, and then decreases monotonically.
Qualitatively the same behavior of $s$ are observed for other potentials
corresponding to different sets of $\{c_n\}$ with $N = 3$ or $N = 5$
(data not shown), which strongly suggests the validity of conjecture (i).

In Fig.~\ref{fig:sym3s}, the analytic results, the leading terms
in  eqs.~(\ref{eq:sw}) and (\ref{eq:sF}), are also plotted.
It is remarkable that the approximate expression (\ref{eq:sw}), which is valid
for small $\beta W$, agrees quite well with the numerical results for $\beta W$
as large as $\beta W \simeq 1$.
For $\beta W$ larger than about unity,
the dependence of $s$ on $F$ is well approximated by the leading term
of eq.~(\ref{eq:sF}) if $F$ is larger than a few to several times 
the maximum slope $V_\text{max}' = \max_x\{V'(x)\}$ of potential $V(x)$; 
in the example shown in Fig.~\ref{fig:sym3s},
$V_\text{max}' \simeq 8.1 W/l$.

In addition to the numerical analysis concerning the dependence of $s$ on $F$,
shown in Fig.~\ref{fig:sym3s}, for more than ten different potentials,
we have carried out more extensive search for possibility of negative $s$.
Symmetric potentials expressed by eq.~(\ref{eq:vsym}) with $N = 3$ and 
those  with $N = 5$ are studied.
For a given $N$, every coefficient $c_n$ ($n = 1, 2, \dots, N$) is 
chosen from a random number uniformly distributed in interval $(-1, 1)$.
For each set $\{c_n\}$ of the coefficients, the potential height $W$ is
chosen from a uniform random number in interval $(0, W_\text{max})$ where
$W_\text{max}$ is set to be $20k_\text{B}T$;
and for a given $W$, the external force $F$ is chosen from a uniform random number
in $(0, F_\text{max}(W))$ where $F_\text{max}$ is set to be $2V_\text{max}'$. 
We have examined 1000 sets of $\{c_n\}$ and 300 sets of $\{W,F\}$ for each
set of $\{c_n\}$ in the case of  $N = 3$, 
and 2000 sets of $\{c_n\}$ and 100 sets of $\{W,F\}$ in
the case of $N = 5$.
In the data of these $5 \times 10^{5}$ samples
we have not detected any instance in which $s < 0$.

All these analytical and numerical investigations firmly indicate that the statement of conjecture (i) should be true.

\subsection{Asymmetric potentials}
\label{sec:asymmetric}

Now we discuss conjecture (ii) associated with asymmetric potentials.
If the potential height is small ($\beta W \ll 1$), 
the analyses of \S\ref{sec:smallF} and \S\ref{sec:smallW} show that
inequality $D > \mu k_\text{B}T$ holds for almost entire range of $F$
except a small interval of order $W/l$.
This interval is given by $0 < F < F_0$ or $F_0 < F < 0$ depending on 
the sign of $F_0$ given by eq.~(\ref{eq:F01}).

If the potential height is not small,
we do not have enough evidence to support conjecture (ii).
It is true that $D - \mu k_\text{B}T$ change its sign at $F = 0$ when $F$ is varied
(\S\ref{sec:smallF}) and that it is positive for large enough $|F|$ (\S\ref{sec:largeF}).
Furthermore, it is shown (\S\ref{sec:smallF}) that in the case of  large potential height 
($\beta W \gg 1$) we have $D < \mu k_\text{B}T$ for $F > 0$ ($F < 0$)
if $a > l/2$ ($a < l/2$) and $|F|$ is small.
These results are consistent with conjecture (ii),
but we are not certain, from the analytical study given above,
whether there is only one interval on the $F$ axis (as the conjecture states)
where inequality $D \ge \mu k_\text{B}T$ is not satisfied.
The numerical investigation presented in \S\ref{sec:example} for 
potential given by eq.~(\ref{eq:V1}) and a similar one (data not shown)
for the piecewise-linear potential (\ref{eq:V3}) support the validity
of conjecture (ii).

\section{Conclusion}
\label{sec:conclusion}

We have postulated two conjectures (\S\ref{sec:conjectures}) concerning
the diffusion coefficient and the differential mobility of a Brownian particle moving
in a one-dimensional periodic potential under the influence of a uniform
external force.
We are quite certain about the validity of conjecture (i) associated with 
symmetric potentials (\S\ref{sec:symmetric}).
It should be possible to prove it mathematically,
although we have not  yet succeeded.
Conjecture (ii) related with asymmetric potentials is partly speculative 
(\S\ref{sec:asymmetric}).

The ratio $\Theta = D/\mu k_\text{B}$ may interpreted as an 
effective temperature\cite{harada04, hayashi04} of the system in 
nonequilibrium steady state.
Then, our conjectures imply that the effective temperature is higher
than the temperature of the heat bath if the potential is symmetric
or if the external force is not too small in the case of asymmetric potential.

Very recently Hayashi and Sasa\cite{hayashi05} have reported an alternative
inequality associated with the diffusion coefficient and the differential mobility.
They have proved that inequality $D/D_0 \ge (\mu k_\text{B}T/D_0)^2$
holds in general for the system considered in the present work.

\section*{Acknowledgement}
The authors would like to thank F.~Matsubara, T.~Nakamura, K.~Hayashi, S.~Sasa and T.~Harada 
for useful comments and discussions.

\appendix
\section{Derivation of eq.~(\ref{eq:Dm})}
\label{sec:appendix}

Our derivation of formula~(\ref{eq:Dm}) is based on 
a prescription to calculate the diffusion coefficient from the solution to
the Fokker-Planck equation.\cite{festa78, risken89, sasaki03}
Let $P(x)$ be the probability distribution function of the particle
in the steady state.
It satisfies the Fokker-Planck equation
\begin{equation}
\label{eq:FP}
	D_0\frac{\d}{\d x}\left(\frac{\d}{\d x} + \beta\frac{\d U}{\d x}\right)P(x) = 0
\end{equation}
for the steady state.
We assume that $P(x)$ is periodic [$P(x + l) = P(x)$] and normalized 
such that $\int_0^l P(x)\,\d x = 1$.
Such a solution is found to be given by
\begin{equation}
\label{eq:Px}
	P(x) = J_{-}(x)/l.
\end{equation}
The average velocity $v$ can be calculated from $P(x)$ as 
\begin{equation}
\label{eq:vp}
	v = -lD_0\left(\frac{\d}{\d x} + \beta\frac{\d U}{\d x}\right)P(x),
\end{equation}
and this leads to formula (\ref{eq:Strat}).
Note that the right-hand side in eq.~(\ref{eq:vp}) is independent of $x$
due to the Fokker-Planck equation (\ref{eq:FP}).
In order to calculate the diffusion coefficient $D$, 
we need to solve the differential equation
\begin{equation}
\label{eq:Qeq}
	D_0\frac{\d}{\d x}\left(\frac{\d}{\d x} + \beta\frac{\d U}{\d x}\right)Q(x) 
	= \left(v + D_0\frac{\d}{\d x}\right)P(x) - \frac{v}{l}
\end{equation}
for $Q(x)$, where $P(x)$ is the probability distribution function given by eq.~(\ref{eq:Px}).
The diffusion coefficient is calculated from a periodic solution 
$Q(x) = Q(x + l)$ to eq.~(\ref{eq:Qeq}) as
\begin{equation}
\label{eq:Dint}
	D = D_0 - \int_0^l \left(\beta D_0\frac{\d U}{\d x} + v\right)Q(x)\,\d x.
\end{equation}
Any periodic solution $Q$ yields the same result for $D$.
Festa and d'Agliano\cite{festa78} solved eq.~(\ref{eq:Qeq}) in the case of
no external force ($F = 0$),  and obtained a formula for $D$,
which is similar to eq.~(\ref{eq:Reim}) but much simpler.
Here, we solve eq.~(\ref{eq:Qeq})  in the case of  nonzero external force,
and derive eq.~(\ref{eq:Dm}).

Integrating eq.~(\ref{eq:Qeq}) once, we have
\begin{equation}
\label{eq:Q1}
	\left(\frac{\d}{\d x} + \beta\frac{\d U}{\d x}\right)Q(x)  = q(x),
\end{equation}
where $q(x)$ is given by
\begin{equation}
\label{eq:qdef}
	q(x) = J_{-}(x)/l + vK_{-}(x)/D_0.
\end{equation}
Here, $K_{-}(x)$ is defined in eq.~(\ref{eq:Kdef}).
We have chosen the integration constant arbitrarily
to get $q(x)$ in eq.~(\ref{eq:Q1}),
since any periodic solution $Q(x)$ is acceptable as remarked above.
Integrating eq.~(\ref{eq:Q1}), we arrive at
\begin{equation}
\label{eq:Qx}
	Q(x) = -\frac{\text{e}^{-\beta U(x)}}{1 - \text{e}^{-\beta Fl}}
		\int_0^l\text{e}^{\beta U(x + y)}q(x + y)\,\d y
\end{equation}
after some manipulations.
This time, the integration constant has been determined such that $Q(x)$
is periodic.

Now we substitute eq.~(\ref{eq:Qx}) into eq.~(\ref{eq:Dint}) to study
the diffusion coefficient.
Making use of eq.~(\ref{eq:Q1}) and the periodicity of $Q(x)$,
we rewrite eq.~(\ref{eq:Dint}) as
\begin{equation}
\label{eq:D1}
	D =D_0 - D_0\int_0^l q(x)\,\d x - v\int_0^l Q(x)\,\d x.
\end{equation}
The second term, without the minus sign, on the right-hand side in this
equation reads
\begin{equation}
\label{eq:Dq}
	D_0\int_0^l q(x)\,\d x = D_0 +vl\langle{K_{-}}\rangle,
\end{equation}
according to the definitions of $q(x)$ and $J_{-}(x)$.
Insertion of eq.~(\ref{eq:Qx}) into the third term in eq.(\ref{eq:D1}) yields the integral
\begin{equation}
	\int_0^l \! \d x\,\text{e}^{-\beta U(x)} \!\!\int_0^l \! \d y\, \text{e}^{\beta U(x+y)}q(x + y)
	= l\!\!\int_0^l \! I_{+}(x)q(x)\,\d x,
\end{equation}
where the right-hand side is obtained by interchanging the order of 
integral and by using the fact
that $U(x+y) - U(x)$ and $q(x)$ are periodic functions of $x$.
From this identity and eqs.~(\ref{eq:Qx}) and (\ref{eq:Strat}) we have
\begin{align}
	v\int_0^l Q(x)\,\d x &= -D_0\int_0^l J_{+}(x)q(x)\,\d x
	\nonumber\\
\label{eq:vQ}
	&= -\mu k_\text{B}T - vl\langle J_{+}K_{-} \rangle,
\end{align}
where the second equality is due to eqs.~(\ref{eq:qdef}) and (\ref{eq:mob}).
Substitution of eqs.~(\ref{eq:Dq}) and (\ref{eq:vQ}) into eq.~(\ref{eq:D1}) gives
eq.~(\ref{eq:Dm}).

The equivalence between formula (\ref{eq:Dm}) for the diffusion coefficient
and the one, eq.~(\ref{eq:Reim}), obtained by other authors can been shown as follows.
Since it can be seen by integration by parts that $\langle{(J_{+} - 1)K_{+}}\rangle = 0$,
eq.~(\ref{eq:Dm}) may be written as
\begin{equation}
	D = \mu k_BT + vl\langle{(J_{+} - 1)(K_{-} - K_{+})}\rangle.
\end{equation}
Now, it is not difficult to see from the definitions of $J_\pm(x)$ that
\begin{equation}
	\frac{\d}{\d x}[J_{-}(x)J_{+}(x)] 
	= \frac{(1 - \text{e}^{-\beta Fl})[J_{-}(x) - J_{+}(x)]}{l\langle{I_\pm}\rangle}.
\end{equation}
This relation and the definition (\ref{eq:Kdef}) of $K_\pm(x)$ lead to
\begin{equation}
	K_{-}(x) - K_{+}(x) 
	= \frac{[J_{+}(x)J_{-}(x) - J_{+}(0)J_{-}(0)]\langle{I_\pm}\rangle}
	{1 - \text{e}^{-\beta Fl}}.
\end{equation}
Substituting this equation into eq.~(\ref{eq:Dm}) and using formula (\ref{eq:Strat})
for $v$, we find 
\begin{equation}
\label{eq:D2}
	D = \mu k_BT + D_0\langle{J_{+}^2 J_{-}}\rangle
		- D_0\langle{J_{+}J_{-}}\rangle.
\end{equation}	
Here, the first and the third terms on the right-hand side cancel out
due to eq.~(\ref{eq:mob}).
Therefore  eq.~(\ref{eq:D2}) is identical to formula (\ref{eq:Reim}),
and the equivalence between eqs.~(\ref{eq:Dm}) and (\ref{eq:Reim}) 
has been proved.


\begin{thebibliography}{99}

\bibitem{kubo91}
R. Kubo, M. Toda and N. Hashitsume:
\textit{Satatistical Physics II: Nonequilibrium Statistical Mechanics\/}
(Springer-Verlag, Berlin, 1991) 2nd ed.

\bibitem{zwanzig01}
R. Zwanzig:
\textit{Nonequilibrium Statistical Mechanics\/}
(Oxford University Press, New York, 2001).

\bibitem{reimann01}
P. Reimann, C. Van den Broeck, H. Linke, P. H\"anggi, J.M. Rubi and A. P\'erez-Madrid:
Phys. Rev. Lett. \textbf{87} (2001) 010602;
Phys. Rev. E\textbf{65} (2002) 031104.

\bibitem{sasaki03}
K. Sasaki: J. Phys. Soc. Jpn. \textbf{72} (2003) 2497.

\bibitem{harada04}
T. Harada and K. Yoshikawa: Phys. Rev. E \textbf{69} (2004)
021113.

\bibitem{hayashi04}
K. Hayashi and S. Sasa: Phys. Rev. E \textbf{69} (2004)
066119.

\bibitem{strat58}
R.L. Stratonovich: Radiotekhinka; elektronika \textbf{3}
(1958) 497, as quoted by Refs.\citen{risken89} and 
\citen{reimann01}.

\bibitem{risken89}
H.~Risken:
\textit{The Fokker-Planck Equation: Methods of Solutions and Applications\/}
(Springer-verlag, Berlin, 1989) 2nd ed. 

\bibitem{festa78} 
R.~Festa and E.~G. d'Agliano: Physica \textbf{90A} (1978) 229.

\bibitem{hayashi05}
K. Hayashi and S. Sasa: cond-mat/0409537 (to be published in Phys. Rev. E).





\end{thebibliography}
\end{document}